\documentclass[graybox]{svmult}

\usepackage{mathptmx}       % selects Times Roman as basic font
\usepackage{helvet}         % selects Helvetica as sans-serif font
\usepackage{courier}        % selects Courier as typewriter font
\usepackage{type1cm}        % activate if the above 3 fonts are
\usepackage{makeidx}         % allows index generation
\usepackage{graphicx}        % standard LaTeX graphics tool
\usepackage{multicol}        % used for the two-column index
\usepackage[bottom]{footmisc}% places footnotes at page bottom

\usepackage[table]{xcolor}

\usepackage{epstopdf}
\usepackage{subfig}
\usepackage{float} 
\usepackage{url}
\usepackage{amsmath,amsfonts,amssymb,amsxtra} 
\newcommand{\NZ}{\mathbb{N}}
\newcommand{\RZ}{\mathbb{R}}
\newcommand{\mbr}{\mathbb{R}}

\newcommand{\mtc}{\mathcal}
\newcommand{\mbf}{\mathbf}

\newcommand{\ind}[1]{{\mbf{1}\{#1\}}}
\providecommand{\abs}[1]{\left\lvert #1 \right\rvert}
\newcommand{\inner}[1]{\left\langle#1\right\rangle}
\newcommand{\scal}[2]{\left\langle #1,#2 \right\rangle}
\providecommand{\norm}[1]{\left\lVert #1 \right\rVert}

\newcommand{\set}[1]{\left\{#1\right\}}

\newcommand{\NV}{\mathcal{N}}  

\newcommand{\probzero}{\ensuremath{\mathbf{P}_0}}
\newcommand{\ew}[1]{\ensuremath{\mathbf{E}\left(#1\right)}}						%E(#1)
\newcommand{\ewt}[2][\theta]{\ensuremath{\mathbf{E}_{#1}\left(#2\right)}}		%E_\theta(#2)

\newcommand{\var}[1]{\ensuremath{\mathbf{Var}\left(#1\right)}}					%Var(#1)
\newcommand{\eps}{\varepsilon}

\newcommand{\manifold}[1]{\mathcal{#1}}
\newcommand{\frames}[1]{\mathcal{#1}}

\newcommand{\EinM}[1]{\mathbf{I}_{#1}}

\newcommand{\diag}[1]{\mathop{\mathrm{diag}\left(#1\right)}}
\newcommand{\supp}{\mathop{\mathrm{supp}}}
\newcommand{\data}[1]{\mathfrak{#1}}   %\data{D}

\newcommand{\dataC}[3]{\mathfrak{#1}=\{\ensuremath{#2_{1},\ldots,#2_{#3}}\}}

\newcommand{\sgn}{\mathop{\mathrm{sgn}}}

\def\cC{{\mtc{C}}}
\newcommand{\wh}[1]{{\widehat{#1}}}

\makeindex             % used for the subject index
\begin{document}

\title*{Construction of Tight Frames on Graphs and Application to Denoising}
%\author{Name of First Author and Name of Second Author}
\author{Franziska G\"obel, Gilles Blanchard, Ulrike von Luxburg}

% \institute{Name of First Author \at Name, Address of Institute, \email{name@email.address}
% \and Name of Second Author \at Name, Address of Institute \email{name@email.address}}

\institute{
Franziska G\"obel \at Institute of Mathematics, University of Potsdam, Germany, \email{goebel@uni-potsdam.de}
\and Gilles Blanchard \at Institute of Mathematics, University of Potsdam, Germany, \email{blanchard@uni-potsdam.de}
%\and Ulrike von Luxburg \at Department of Computer Science, University of Hamburg, Germany, \email{luxburg@informatik.uni-hamburg.de}
\and Ulrike von Luxburg \at Department of Computer Science, University of T\"ubingen,\email{luxburg@informatik.uni-tuebingen.de}
}
\maketitle

\abstract{Given a neighborhood graph representation of a finite set of points $x_i\in\RZ^d,i=1,\ldots,n,$  we construct 
a frame (redundant dictionary) for the space of real-valued %square-integrable 
functions defined on the graph. This frame is adapted to the underlying
geometrical structure of the $x_i$, has finitely many elements, and these elements are localized in 
frequency as well as in space. This construction follows the ideas
of \cite{Hammond11}, with the key point that we
construct a tight (or Parseval) frame. 
This means we have a very simple, explicit reconstruction formula
for every function $f$ defined on the graph from the coefficients given by 
its scalar product with the frame elements.
We use this representation in the setting of denoising where we are given noisy observations of a function $f$ defined on the graph. By applying a thresholding method to the coefficients in the reconstruction formula, we define an estimate of $f$ whose risk satisfies a tight oracle inequality.}

\section{Introduction}

\subsection{Motivation}

When dealing with high-dimensional data, a general principle is that 
the curse of dimensionality can be efficiently fought if one assumes the data points
to lie on a structure of smaller intrinsic dimensionality, typically a manifold. Some well-known 
methods to discover %and take advantage of 
such a lower dimensional structure include
Isomap \cite{isomap00}, LLE \cite{lle00} and Laplacian Eigenmaps \cite{lap_eig03}.

In this work, our main interest is not in visualizing or representing by an explicit mapping
the underlying structure of the observed data points; rather, we want
to represent or estimate efficiently a real-valued function on these points. More specifically,
we focus on the following {\em denoising} problem: assuming we observe a noisy
version of the function $f$, $y_i=f(x_i) + \eps_i$ at  points $(x_1,\ldots,x_n)$,
we would like to recover the values of $f$ at these points. An important step
for solving this problem is to find a dictionary of functions to represent the signal
$f$, which is adapted to the structure of the data. Ideally, we would like
this dictionary to exhibit the features of a wavelet basis. 
In traditional signal processing on a flat space, with data points on a regular grid, 
orthogonal wavelet bases offer a very powerful tool to  sparsely represent
signals with inhomogeneous regularity (such as a signal that is very smooth
everywhere except at a few singular points where it is discontinuous).
Such bases are in particular well suited to the denoising task. Can this be generalized
to irregularly scattered data on a manifold?

We present such a method to construct a so-called {\em Parseval frame}
of functions exhibiting wavelet-like properties while adapting to the intrinsic
geometry of the data. Furthermore, we use this dictionary
for the denoising task using a simple coefficient thresholding method.

This work is organized as follows. In the coming section, we discuss the
relationship to previous work on which the present paper is built, as well
as pointing out our new contributions. In Section 2, we recall important notions
of frame theory as well as of neighborhood graphs needed for our construction.
The construction of the frame and its properties is presented in Section 3.
In Section 4, we develop a coefficient thresholding strategy for the denoising problem.
In Section 5, we present numerical results and method comparison on testbed data.

\subsection{Relation to previous work}

Regression methods that adapt to an underlying lower dimension of the data have been
considered by \cite{BiLi,KpoDas,Kpo} using local polynomial estimates, 
random projection trees, and nearest-neighbors, respectively. However, these
methods are not constructed to adapt to an inhomogeneous regularity of the
target function: in these three cases, the smoothing scale (determined by
the smoothing kernel bandwidth, the tree partition's average data diameter,
or the number of neighbors, respectively) is fixed globally. In the experimental
section \ref{sec:exp}, for data lying on a smooth manifold but a target function
exhibiting a sharp discontinuity, we demonstrate the advantage of our method
over kernel smoothing.

Based on motivations similar to ours, 
a method for constructing a wavelet-like basis on scattered data 
was proposed by \cite{Gavish10}. 
It is based on a hierarchical tree partition of the data, on which 
a Haar-like basis of 0-1 functions is constructed. However, 
the performance of that method is then adapted to the geometry of the {\em tree},
in the sense that the distance of two points is measured through tree path distance.
This can strongly distort the original distance: two close points in original distance
can find themselves in very separated subtrees.

The construction proposed here, based on a transform of the spectral decomposition
of the graph Laplacian, follows closely the ideas of \cite{Hammond11}.
Two important contributions brought forth in the present work are that we construct
a Parseval (or tight) frame, rather than a general frame; and we consider an
explicit thresholding method for the denoising problem. The former point is 
crucial to obtain sharp bounds for the thresholding method, and also 
eliminates the computational problem of signal reconstruction from the frame coefficients,
since Parseval frames enjoy a reconstruction formula similar to that of an orthonormal basis.
The choice of multiscale bandpass filter functions leading to the tight frame
is inspired by the recent work of \cite{gerard}, where the spectral decomposition principle
is also studied, albeit in the setting of a quite general metric space.

\section{Notations - Basics}
\subsection{Setting}

We consider a sample of $n$ points $x_i\in\RZ^d$. These points are assumed to belong to an unknown low-dimensional submanifold 
$\manifold{M}\subset\RZ^d$. We denote the design by $\dataC{D}{x}{n}\subset\manifold{M}$. 
Furthermore, we observe on these points the (noisy) value of a function $f: \data{D} \rightarrow \RZ$. Since $\data{D}$ is finite, we can represent the function $f$ as vector $f=(f(x_1),\ldots,f(x_n))^t\in\RZ^n$. The space of all (square-integrable) functions $f$ defined on $\data{D}$ 
 is denoted $L^2(\data{D})$ and endowed with the usual Euclidean inner product.

We denote by $y_i=f(x_i)+\epsilon_i$ the noisy observation of $f$ at $x_i$, where $\epsilon_i$ are independent identically distributed centered random variables. The problem we consider in this work
is that of  denoising, that is, try to recover the
underlying value of the function $f$ at the points $x_i$.

While the existence of a low-dimensional supporting manifold $\manifold{M}$ for the design points motivates the construction of the proposed method, we underline (again) that $\manifold{M}$ is not known to the user and the method only uses the knowledge of the design points.
In such a setting, a key idea to recover implicitely some information on the geometry of $\manifold{M}$ is to construct a neighborhood graph based on the design points (see Section \ref{sec:graph} for details).

\subsection{Frames}
\label{sec:frames}

For the construction in Section \ref{sec:const}, we rely on the notion of
a {\em vector frame}, for which we recall here some important properties
(see e.g. \cite{FF2012},  \cite{Han07} and \cite{Christensen08}). A frame is an overcomplete
dictionary with particular properties allowing it to act almost as basis.

\begin{definition}
Let $\mathcal{H}$ be a Hilbert space. Then a countable set $\{z_i\}_{i\in I}\subset \mathcal{H}$ is a frame with frame bounds $A$ and $B$ for $\mathcal{H}$ if there exists constants $0<A\leq B <\infty$ such that \begin{equation}\label{frameCond}
 \forall z\in\mathcal{H}:~\; A \norm{z}^2 \leq \sum_{i\in I}\abs{\scal{z}{z_i}}^2\leq B \norm{z}^2.
 \end{equation}
A frame is called tight if $A=B$, in particular the frame is called Parseval if $A=B=1$.
\end{definition}

In the remainder of this work we consider the case of a Euclidean space $\mathcal{H}=\RZ^n$, and assume that $\{z_i\}_{i\in I}$ is a
frame with a finite number of elements.
Two important operators associated to the 
frame are the {\em analysis} operator 
\begin{equation}T:\RZ^n \rightarrow \RZ^I,~ Tz:=(\scal{z}{z_i})_{i\in I}\end{equation} (sequence of frame coefficients), and its adjoint
the {\em synthesis} operator:
\begin{equation}T^*: \RZ^I\rightarrow \RZ^n,~ T^*a=T^* (a_i)_{i\in I}^t=\sum_{i\in I} a_i z_i.\end{equation}
Further, the {\em frame} operator is defined as $S=T^*T$: %concatenation of analysis and synthesis operator: 
\begin{equation}S:\RZ^n \rightarrow\mathcal \RZ^n, ~Sz=T^*Tz=\sum_{i\in I}\scal{z}{z_i}z_i,\end{equation} and
finally the {\em Gramian} operator as
$U=TT^*$,
\begin{equation}U: \RZ^I\rightarrow \RZ^I,~ Ua=TT^*a= \left\{\scal{ \sum_{i\in I} a_i z_i}{z_k}\right\}_{k\in I}.\end{equation}
In matrix form, the columns of $T^*$ are the vectors $z_i, i\in I$,
$T$ is its transpose and $U_{ij} = \scal{z_i}{z_j}$.

The definition of a frame implies that $S$ is invertible, and
it is possible to reconstruct any $z$ from its frame coefficients by
$z=\sum_{i\in I}\scal{z}{z_i}z^*_i=\sum_{i\in I}\scal{z}{z^*_i}z_i$,
where $z^*_i:=S^{-1}z_i, i\in I$ is called the {\em canonical dual} frame of $(z_i)_{i \in I}$.

We recall some properties of finite Parseval frames 
over Euclidean spaces (see e.g. \cite[chapter 3]{Han07} ).

\begin{theorem}[Properties of Parseval frames]\label{t:pars}
Let $\mathcal{H}$ be a Hilbert space with $\dim \mathcal{H}=n< \infty$. The following statements are equivalent:
\begin{enumerate}
\item  	$\{z_i\}_{1\leq i \leq k}\subset \mathcal{H}$ is a Parseval frame.
\item 	$\forall y\in\mathcal{H}: ~~y=\sum_{i=1}^{k}\scal{y}{z_i}z_i$  % (reconstruction formula)
\item the frame operator $S$ is the identity on $\RZ^n$.
\item the Gramian operator $U$ is an orthogonal projector of rank $n$
in $\RZ^k$.
\end{enumerate}
Furthermore if $\{z_i\}_{1..k}\subset \mathcal{H}$ is a Parseval frame, then 
\begin{itemize}
\item  $\norm{z_i}\leq 1$ for $i\in \{1,\ldots,k\}$.
\item  $\dim \mathcal{H}=n=\sum_{i=1}^{k}\norm{z_i}^2$.
\item the canonical dual frame is the frame itself.
\end{itemize}
\end{theorem}
For the present work, the two most important points of this theory are 
the following: first, the reconstruction formula (point 2 above), where we see that a
Parseval frame acts similarly to an orthonormal basis; secondly, 
if we construct a vector $v=T^*a=\sum_i a_i z_i$ from an arbitrary vector of
coefficients $(a_i)$, then
\begin{equation}
\label{eq:analnorm}
\norm{\sum_i a_i z_i}^2 = \inner{T^*a,T^*a} = \inner{a,Ua} = \norm{Ua}^2 \leq \norm{a}^2,
\end{equation}
which follows from property 4 above.

\subsection{Neighborhood Graphs}
\label{sec:graph}

In order to exploit the structure and geometry of the unknown submanifold $\manifold{M}$ on which
the sample $\data{D}$ is supposed to lie, a powerful idea 
is to use a graph-based representation of the data $\data{D}$ through a {\em neighborhood graph}.
The points in $\data{D}$ correspond to the vertices of the graph,
and two vertices of the graph are joined by an edge when the two corresponding points are
neighbors (in some appropriate sense) in $\RZ^n$. The underlying idea is that the
local geometry of $\RZ^n$ is reflected in the local connectivity of the graph,
while the long-range geometry of the graph reflects the geometrical properties of
the manifold $\manifold{M}$, rather than those of $\RZ^n$.

Formally, a finite graph $G=(V,E)$ is given by a finite set of vertices $V$ )
and a set of edges $E\subset V\times V$.  The $|V|\times|V|$ adjacency matrix $A$ of the 
graph is defined by $A_{i,j}=1$ if $(v_i,v_j)\in E$ and $A_{i,j}=0$ otherwise. An undirected graph is such that its adjacency matrix is symmetric.

The graph is called weighted if every edge $e\in E$ has a positive weight $w(e)\in\RZ_{+}$.
In this case the notion of adjacency matrix is extended to $A_{i,j}=w((v_i,v_j))$ 
if $(v_i,v_j)\in E$ and $A_{i,j}=0$ otherwise.
The degree of a vertex $v_i$ in a (possibly weighted) graph is defined as $d_i=d(i)=\sum_{j=1}^{|V|}A_{i,j}$. 

As announced, we focus on geometric graphs, which (can) approximate the structure 
 of the unknown $\manifold{M}$.
Each point $x_i$ is represented by a vertex, say $v_i$. An edge between two vertices represents a small distance, or a high similarity, of the two associated points. The weight of an edge can quantify the similarity more finely.\\
We use the Euclidean distance $d(x_i,x_j)=\norm{x_i-x_j}.$ 
We recall three usual ways to construct the edges of a neighborhood graph:
\begin{itemize}
\item (undirected) $k$-nearest-neighbor graph: an undirected edge connects the two vertices $v_i$ and $v_j$ iff $x_i$ belongs to the $k$ nearest neighbors of $x_j$, or $x_j$ belongs to the $k$ nearest neighbors of $x_i$ ("the k-NN-graph"). 
\item $\epsilon$-graph: an undirected edge connects two vertices $v_i$ and $v_j$ iff

$d(x_i,x_j)<\epsilon$.
\item complete weighted neighborhood graph: for each pair of vertices there exists an undirected edge with a weight depending on the distance/similarity of the two vertices.
\end{itemize}
A $k$-NN graph or an $\epsilon$-graph can be made weighted by additionally assigning weights to the edges depending on $d(x_i,x_j)$, for instance by choosing Gaussian weights $w(\set{i,j})= \exp ( - d^2(x_i,x_j)/2\lambda^2)$.

\subsection{Spectral Graph Theory}

If one considers real-valued functions $f:\manifold{M}\rightarrow \RZ$ defined on a submanifold $\manifold{M}\subset\RZ^d$, it is known that under some regularity assumptions on the submanifold $\manifold{M}$, the eigenfunctions of the Laplace-Beltrami-operator 
give a basis of the space of squared-integrable functions on $\manifold{M}$.
Since $\manifold{M}$ is unknown in our setting, the principle of the {\em
Laplacian Eigenmaps} method \cite{lap_eig03} is to use 
a discrete analogon, namely the graph Laplace operator $L$ on a neighborhood graph.

Given a finite weighted undirected graph with adjacency matrix $A$ ($n \times n$) and vertex degrees $(d_i)_i$, as introduced in the previous section, we will either use the unnormalized graph Laplace operator  $L^{u}$ or the normalized (symmetric) graph Laplace operator $L^{norm}$
 defined by
\begin{eqnarray}L^{u}&=&D-A \\
 L^{norm}&=&\EinM{n}-D^{-1/2}AD^{-1/2},\nonumber \end{eqnarray}
where $D=\diag {d_1,\ldots, d_n}$ is a diagonal matrix with entries $d_i$ on the diagonal. 
By construction  $L^{u}$ and $L^{norm}$ are symmetric matrices. The positive semidefiniteness follows from   $f^t L^{u}f=0.5 \sum_{(i,j)}A_{i,j} (f_i-f_j)^2 $  and $f^t L^{norm}f=0.5 \sum_{(i,j)}A_{i,j} (\frac{f_i}{\sqrt{d_i}}-\frac{f_j}{\sqrt{d_j}})^2$ respectively.
The spectral theorem for matrices indicates that the normalized eigenvectors $\Phi_i$ of the graph Laplace operator $L$ ($L^{u}$ resp. $L^{norm}$) 
form an orthonormal basis of  $\RZ^n$ and all eigenvalues are nonnegative. Furthermore the number of components of the graph is given by the number of eigenvalues equal to 0.

\begin{figure*}
\centering
\subfloat[]{
\includegraphics[width=0.8\textwidth]{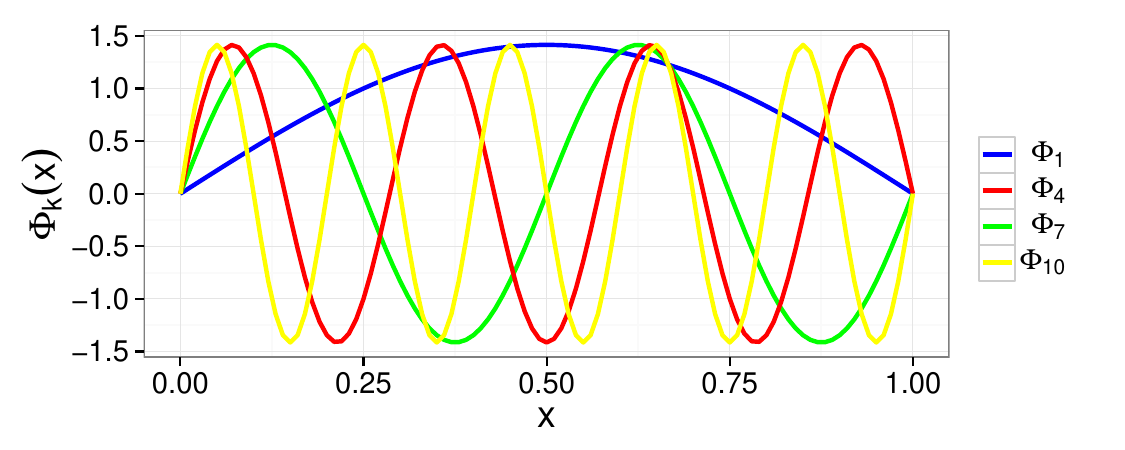}}\hfill
\subfloat[]{  \includegraphics[width=0.8\textwidth]{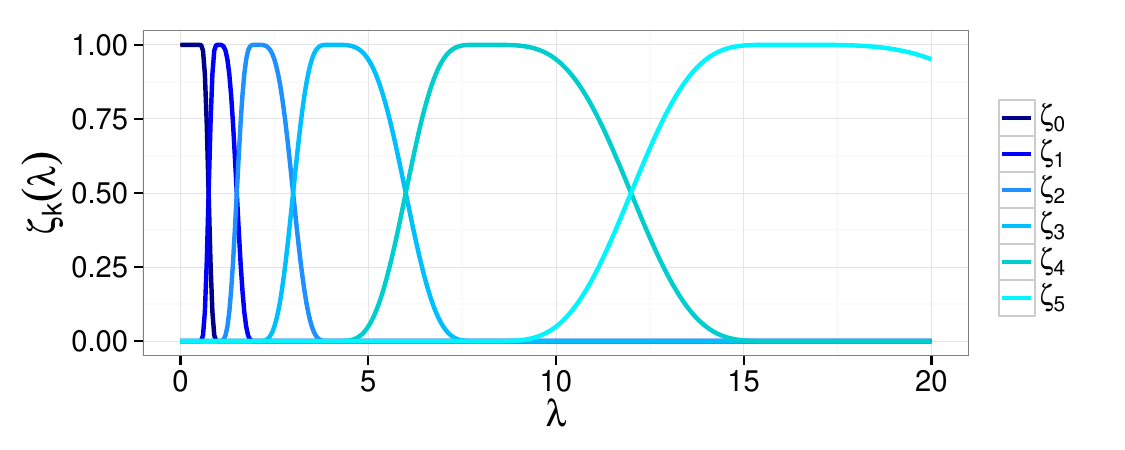}}\hfill
\subfloat[]{ \includegraphics[width=0.8\textwidth]{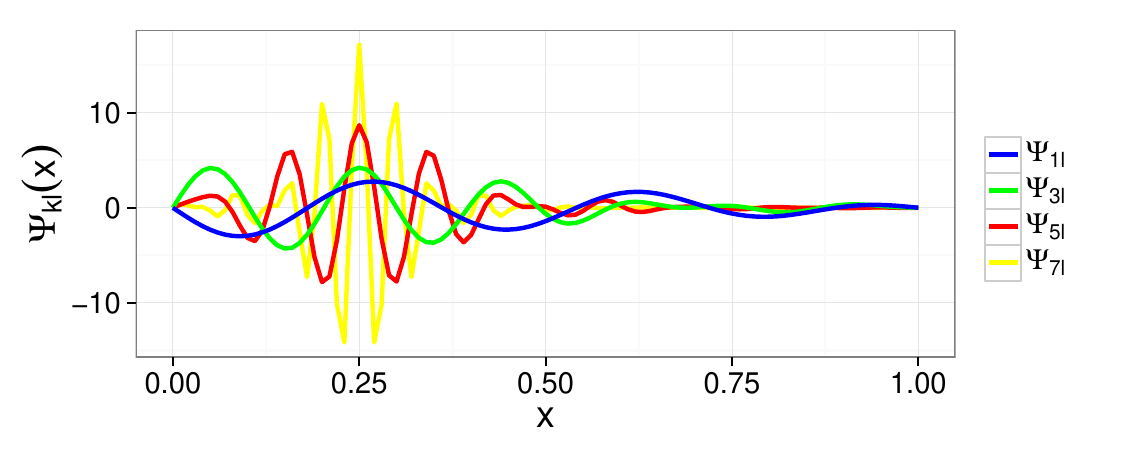}}

  \caption{\label{fig:1} Littlewood-Paley on $L^2(0,1)$: (a) eigenfunctions; (b) multiscale bandpass filter;
(c) frame elements.}
 \end{figure*}

\section{Construction and Properties}
\label{sec:const}
\subsection{Construction of a tight graph frame}

As discussed earlier, the principle of Laplacian Eigenmaps is to use
the basis $(\Phi_i)_{1 \leq i \leq n}$ to represent and process the data.
An important advantage of this basis as compared with the natural basis of $\mbr^d$ 
is that it will be {\em adapted} to the geometry of the underlying submanifold
$\manifold{M}$ supporting the data distribution. For instance, in the denoising
problem, a reasonable estimator of $f$ could be 
a truncated expansion of the noisy vector of observations $Y$ in the basis
$(\Phi_i)_{1 \leq i \leq n}$. 

On the other hand, a disadvantage of this basis is that
it is not {\em spatially localized}. To get an intuitive view, consider the simple
case of the interval $[0,1]$ with uniformly distributed data. In the population view,
the eigenbasis of the Laplacian is the Fourier basis. While a truncated expansion in this basis is well-adapted
to represent functions that are uniformly regular, it is not well-suited for functions
exhibiting locally varying regularity (as an extreme example, 
a signal that is very smooth
everywhere except at a few singular points where it is discontinuous). 
By contrast, wavelet bases, because they are
localized both in space and frequency, allow for an efficient (i.e. sparse)
representation of signals with locally varying regularity.

If we now think of data supported on a one-dimensional submanifold (curve) of $\mbr^d$,
we can expect that the Laplacian eigenmaps method will discover a warped Fourier
basis following the curve; and, for a more general submanifold $\manifold{M}$,
``harmonics'' on $\manifold{M}$.

In order to go from this basis to a spatially localized
dictionary, following ideas of \cite{gerard} and \cite{Hammond11},
we use the principle of the {\em Littlewood-Paley} decomposition.

Let $G$ be an undirected geometric neighborhood graph with adjacency matrix $A$ constructed from $\data{D}$, and $L$ be an associated symmetric graph Laplace operator with increasing eigenvalues $0=\lambda_1\leq \lambda_2\leq\ldots\leq\lambda_n$ and normalized eigenvectors $\Phi_i\in\RZ^n,i=1..n$.

We first define a set of vectors using a decomposition of unity and a splitting operation and we will show that this vector set is a Parseval frame.
\begin{definition}\label{frame}
Let $\{\zeta_k\}_{k\in \NZ}$ be a sequence of functions $\zeta_k: \RZ_{+}\rightarrow [0,1]$ satisfying 
\begin{description}
\item[(DoU)] $~~\sum_{j\geq 0}\zeta_j(x)=1 \label{e:zeta}$ for all $x\geq 0$;
\item[(FD)] $~~\#\{\zeta_k: \zeta_k(\lambda_i)\neq 0\}<\infty$  for $ i=1,\ldots,n$.
\end{description}
Then we define the set of column vectors $\{\Psi_{kl}\in\RZ^n , 0\leq k \leq Q, 1 \leq j \leq n\}$ by
\begin{equation}
\Psi_{kl}=\sum_{i=1}^{n}\sqrt{\zeta_k(\lambda_i)}\Phi_i(x_l) \Phi_i. %\\
\end{equation}
with
$Q:=\max\{k:  \exists i \in \set{1,\ldots,n} \text{ with } \zeta_k(\lambda_i)>0\}$.
\end{definition}

\begin{theorem}\label{satz3} $\{\Psi_{kl}\}_{k,l}$ is a Parseval frame for $\mathcal{H}=\RZ^n$, that is for all $x\in\RZ^n$: 
\begin{equation}
\sum_{k,l}\abs{\scal{x}{\Psi_{kl}}}^2 = \norm{x}^2\,.
\label{Parseval}
\end{equation}
\end{theorem}
\begin{proof}
If we can show that $\sum_{(k,l)}\Psi_{kl}^{} \Psi_{kl}^t=\EinM{n}$,  
 we get immediately  
\begin{equation}
y=\EinM{n} y=\left(\sum_{(k,l)}\Psi_{kl}^{} \Psi_{kl}^t\right) y
\stackrel{}{=}   \sum_{(k,l)} \scal{y}{\Psi_{kl}}\Psi_{kl},
\end{equation}
for $y\in\RZ^n$.
According to theorem \ref{t:pars} this equation is equivalent to the condition (\ref{frameCond}) with $A=B=1$. So we are done.
It remains to show $\sum_{k,l}\Psi_{kl}^{} \Psi_{kl}^t=\EinM{n}$. 
We have (since we sum over a finite number of elements)
\begin{eqnarray}
\sum_{(k,l)} \Psi_{kl}^{} \Psi_{kl}^t 
& = & \sum_{k,l,i,j} \sqrt{\zeta_k(\lambda_i)} \sqrt{\zeta_k(\lambda_{j})} \Phi_i(x_l) \Phi_{j}(x_l) \Phi_i \Phi_{j}^t \nonumber\\
&=&\sum_{i=1}^{n} \sum_{k=0}^{Q} \zeta_k(\lambda_i) \, \Phi_i \Phi_i^t \nonumber\\
& = &\sum_{i=1}^{n} \Phi_i \Phi_i^t=\EinM{n}.
\end{eqnarray}
For the second equality, we have used that $\sum_l  \Phi_i(x_l) \Phi_{j}(x_l) = \inner{\Phi_i,\Phi_j} = \ind{i=j}$,
since $\{\Phi_i\}_i$ is an orthonormal basis (onb). For the third equality, we used
(DoU), and for the last again the onb property.
\qed
\end{proof}

We now choose a special sequence of functions satisfying the decomposition of unity (DoU) 
condition while also ensuring (a) a spectral localization property for the frame elements
and (b) a multiscale decomposition interpretation of the resulting decomposition.
This construction follows \cite{gerard}, and is known in the context of
functional analysis as a smooth Littlewood-Paley decomposition.  
\begin{definition}[Multiscale bandpass filter]\label{bandpass} 
Let $g\in C^{\infty}(\RZ_{+})$, $\supp g \subset [0,1]$, $0\leq g \leq 1$, $g(u)=1$ for $u\in [0,1/b]$ (for some constant $b>1$). For $k\in\NZ=\{0,1,\ldots\}$ the functions $\zeta_k: \RZ_{+}\rightarrow [0,1]$ are defined by
\begin{equation}
\label{defzeta}
\zeta_k(x):= \begin{cases}g(x) & \text{if } k=0\\
g(b^{-k}x)-g(b^{-k+1}x) & \text{if } k>0\\
\end{cases}
\end{equation} 
The sequence $\{\zeta_k\}_{k\geq 0}$ is called multiscale bandpass filter.
\end{definition}
This definition leads to the following properties: $\zeta_k(x)=\zeta_1(b^{-k}x)$ 
for $k \geq 1$ (multiscale decomposition), 
 $\zeta_k\in C^{\infty}(\RZ_{+})$, $0\leq\zeta_k\leq 1$, $\supp \zeta_0\subset[0,1]$, $\supp \zeta_k\subset [b^{k-2},b^{k}]$ for $k\geq 1$ (spectral localization property). Moreover, one can
check readily
\begin{equation}\sum_{j\geq 0}\zeta_j(x)=1,\end{equation} i.e., the (DoU) condition holds.
In practice, we use a dyadic bandpass filter, that is, $b=2$. The functions $\zeta_0,\ldots,\zeta_5$ with $b=2$ are displayed in Figure \ref{fig:1}b. By construction, the parameter $k$ in $\Psi_{kl}$ is naturally a spectral scale
parameter, while $l$ is a spatial localization parameter: the frame element $\Psi_{kl}$ is localized
around the point $x_l$, as we discuss next.

\subsection{Spatial localization}

By construction, the elements of the frame are band-limited, i.e.
localized in the spectral scale, in the sense that for a fixed $k$,
the frame elements $\Psi_{kl}$ ($l=1,\ldots,n$) are linear combinations of
the eigenvectors of the graph Laplacian (``graph harmonics'') corresponding to
eigenvalues in the range $[b^{k-2},b^k]$ only.

From our initial motivations, it is desirable that in
contrast with the eigenfunctions of the Laplace operator, the frame elements $\Psi_{kl}$ are
 spatially localized functions. In the classical Littlewood-Paley construction for the usual Laplacian on
the interval $[0,1]$, this is a well-known fact: the use of linear combination of trigonometric
functions $\Psi_{k l}(y) := \sin(kl)\sin(ky)$ via smooth multiscale bandpass filters weights as described
in Definition \ref{bandpass} gives rise to strongly localized functions (as illustrated
in Figure~\ref{fig:1}). 

\begin{figure*}
\centering 

\hspace*{0.2cm}
 
\includegraphics[width=0.4\textwidth]{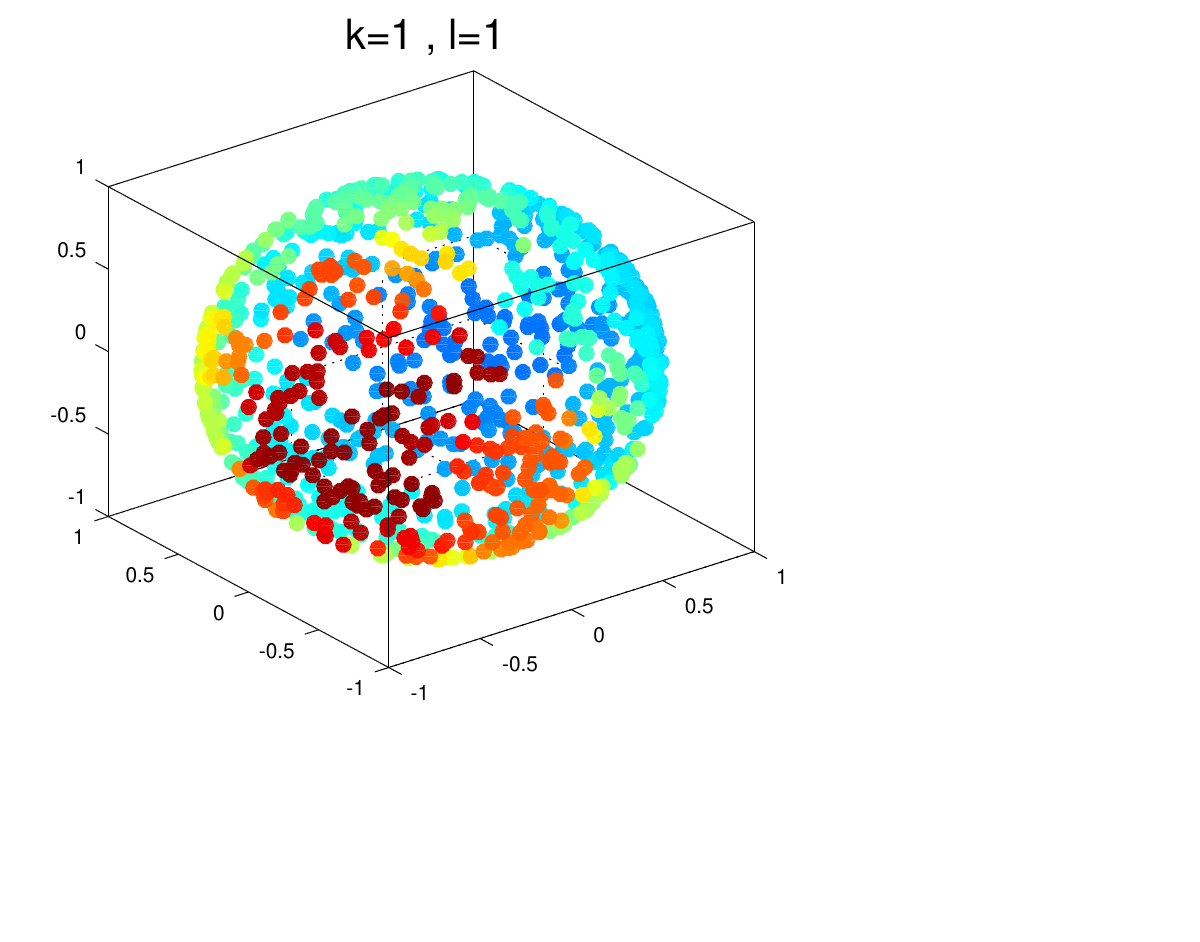} \hspace{-1.4cm}
\includegraphics[width=0.4\textwidth]{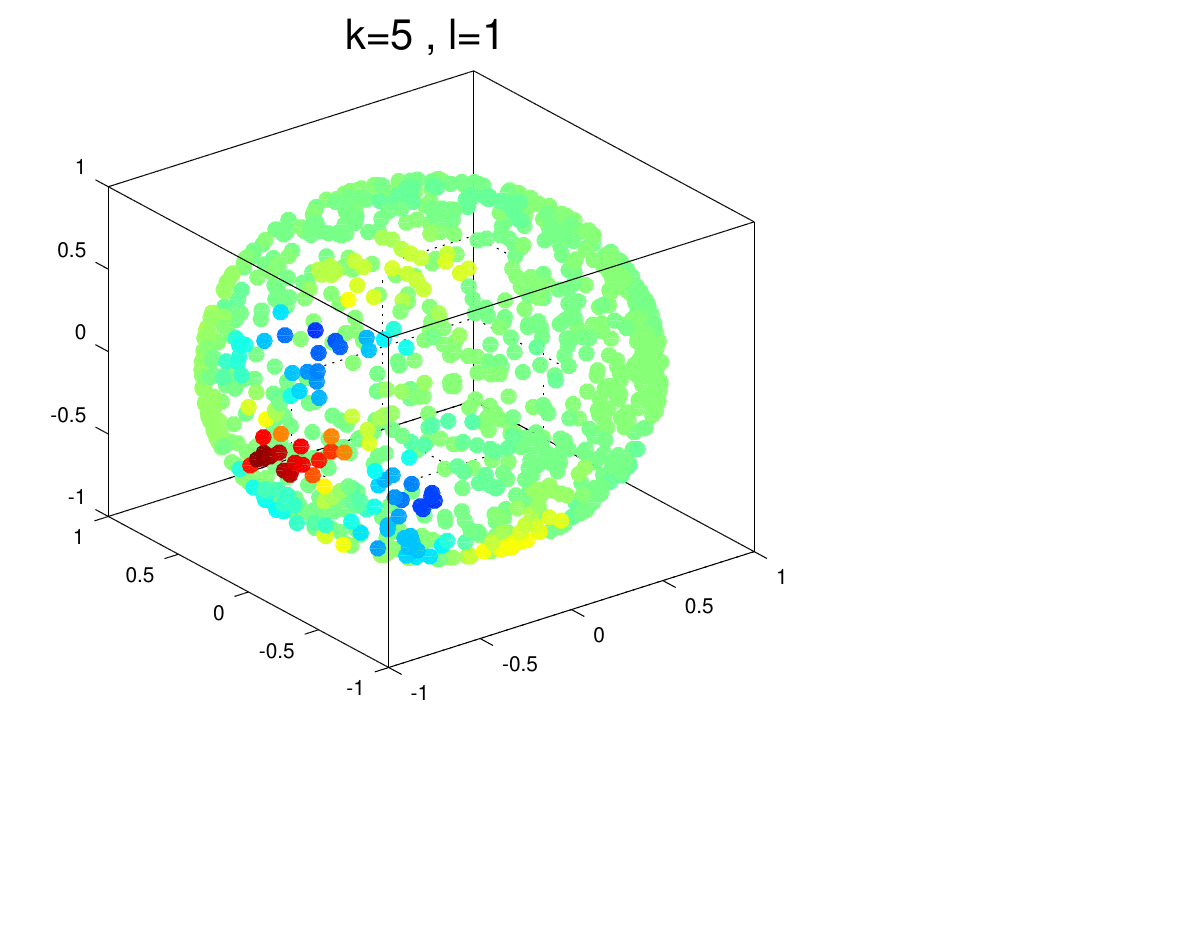} \hspace{-1.4cm}
\includegraphics[width=0.4\textwidth]{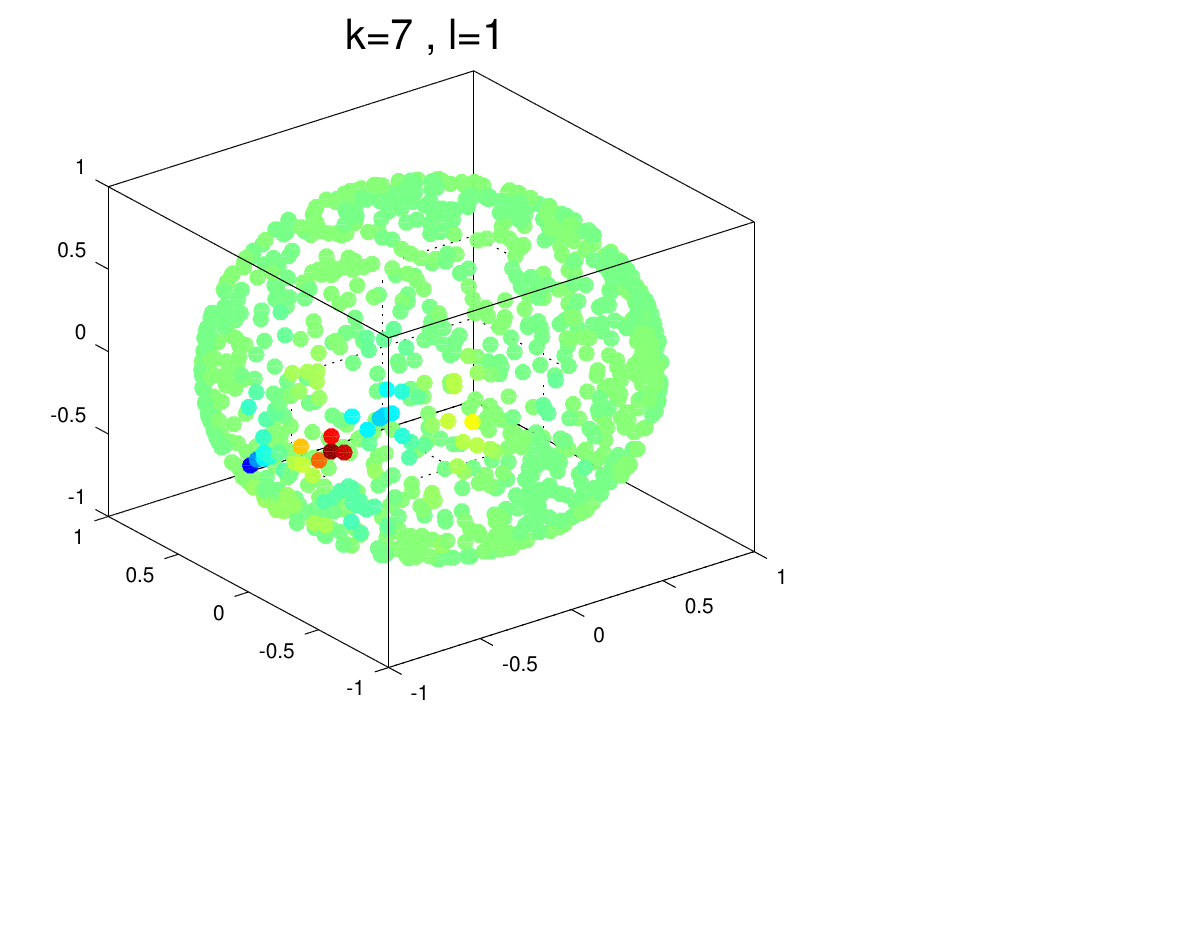} \hfill
\vspace{-1cm}
\caption{Sphere data: frame elements $\Psi_{kl}$ for $l$ fixed and $k=1,5,7$}
\label{fig:locSphere}
\end{figure*}

Regarding the corresponding discrete construction based on the graph Laplacian,
this localization property is certainly observed in practice (as illustrated in Figure \ref{fig:locSwiss} and \ref{fig:locSphere}, see Section \ref{sec:exp} for the setup of the numerical experiments).

Concerning the theoretical perspective, we first review briefly the existing results of \cite{Hammond11}, denote $d$ the shortest path distance in the graph. Theorem 5.5 of \cite{Hammond11} there gives the following localization result  for graph frames:
\begin{equation}
\label{vdg:conc}
\frac{\Psi_{kl}(x)}{\norm{\Psi_{kl}}_2} \leq C b^{-k}\,,
\end{equation}
for all $x$ with $d(x,x_l) \geq K$,  
under the assumption that the scaling function $\zeta_1$ is $K$-times differentiable with vanishing first $(K-1)$  derivatives
in 0, non-vanishing $K$-th derivative, and the scale parameter $k$ is big enough. 
This says that $\Psi_{kl}$ is ``localized'' around the point $x_l$.
Unfortunately, this result is not informative
in our framework for two reasons: first, we chose a function $\zeta_1$ (see \eqref{defzeta}) vanishing
in a neighborhood of zero, so that all derivatives vanish in the origin, contradicting one of the above
assumptions. Secondly, and independently of this first issue, the condition ``$k$ is big enough'',
as well as the factor $C$, depend on the size $n$ of the graph and of the largest eigenvalue of the Laplacian. As a consequence it is unclear if this bound covers any interesting part of the spectrum 
(for $k$ too large, the spectral support $[b^{k-2}, b^k]$ does not contain any eigenvalues, so that $\Psi_{kl}$
is trivial). Finally, for fixed $k$ 
the bound also does not give information on
the behavior of $\Psi_{kl}(x)$ when the path distance of $x$ to $x_l$ becomes very large.

On the other hand the form of the scaling function $\zeta_1$ used in the present work  is based on \cite{gerard}
where a theory of multiscale frame analysis is developed on very general metric spaces
under certain geometrical assumptions. Without entering into detail, it is proved there
that using this construction, the obtained frame functions $\Psi_{kl}(x)$
are upper bounded by  $O( (d(x,x_l)/b^k)^{-\nu})$ for $\nu$ arbitrary large.
We observe that this type of localization estimate
is sharper than \eqref{vdg:conc} for fixed $x$ and growing $k$, as well as for fixed scale $k$ and varying $x$.
We conjecture that these theoretical results apply meaningfully in the discrete setting considered here, under the assumption that $x_1,\ldots,x_n$ are iid from a sufficiently regular distribution $\probzero$ on a regular manifold $\manifold{M}$,
but it is out of the intended scope of the present paper 
to establish this formally. 
In particular ``meaningfully'' means that the constants involved in the bounds should be independent of the 
graph size (otherwise
the bounds could potentially be devoid of interest for any particular graph, as
pointed out above), a question that we are currently investigating.

\newcommand{\QFigScalqq}{0.1}
\newcommand{\Grescalq}{0.21}

\begin{figure*}
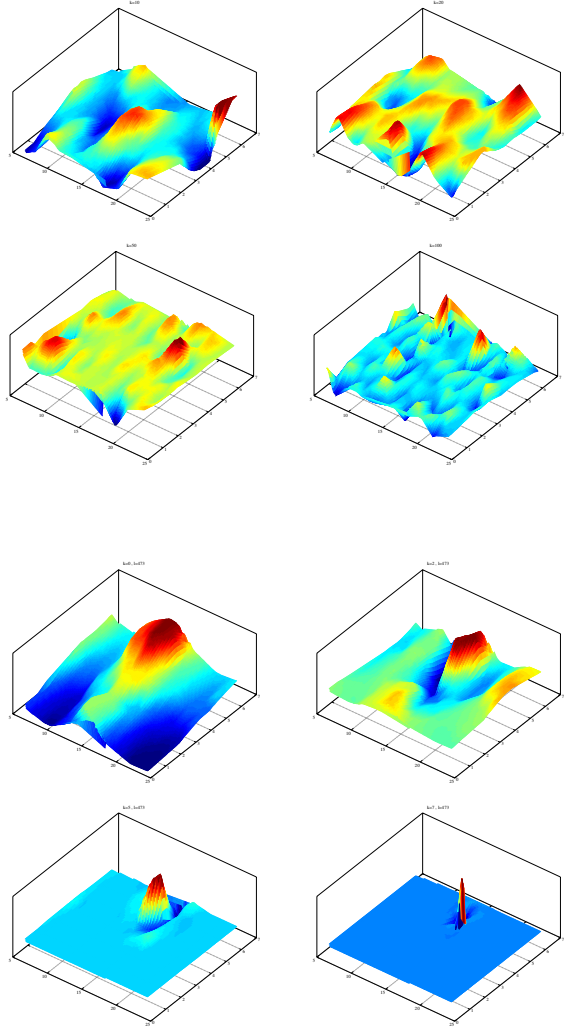

\centering

\scalebox{\Grescalq}{\input{GGG010}}\hspace{-3mm}
\scalebox{\Grescalq}{\input{GGG020}}\hspace{-3mm}
\scalebox{\Grescalq}{\input{GGG050}}\hspace{-3mm}
\scalebox{\Grescalq}{\input{GGG100}}
\vspace{1cm}

\scalebox{\Grescalq}{\input{G00}}\hspace{-3mm}
\scalebox{\Grescalq}{\input{G02}}\hspace{-3mm}
\scalebox{\Grescalq}{\input{G05}}\hspace{-3mm}
\scalebox{\Grescalq}{\input{G07}} 
  \caption{Swiss roll data:  top: eigenvectors $\Phi_j$ for $j=10, 30, 50, 100$; bottom:
frame elements $\Psi_{kl}$ for $l$ fixed and $k=0,2,5,7$.
(Construction from actual swiss roll data,  then 
``unrolled'' for clearer graphical representation.)}
\label{fig:locSwiss}
\end{figure*}

\section{Denoising}

We consider the regression model for fixed design points $\data{D}=\set{x_i, i=1..n}$ 
and observations $y_i=f(x_i)+\epsilon_i$ ($\epsilon_i$ are independent and identically distributed random variables with $\ew{\epsilon_i}=0$ and $\var\eps_i=\sigma^2$). The aim of denoising is to recover the function $f:\data{D}\rightarrow\RZ$ at the design points themselves. We will use the proposed Parseval frame in order to define an estimate $\wh{f}$ of the function $f$.
In what follows, since the $\data{D}$ is fixed, we identify $f$ with the vector
$(f(x_1),\ldots,f(x_n))$ and denote $y=(y_1,\ldots,y_n)$.
 
Given the frame $\frames{F}$ with a multiscale bandpass filter  as defined in \ref{frame} and \ref{bandpass} associated to the data points $\data{D}$,
we denote the frame coefficients
$a_{kl}=\scal{\Psi_{kl}}{f}$ for $f$ and $b_{kl}=\scal{\Psi_{kl}}{y}$ for $y$. Due to the linearity of the inner product we get
$a_{kl}=b_{kl}-\scal{\Psi_{kl}}{\epsilon}.$
We estimate the unknown coefficients $a_{kl}$ by adjusting the known coefficients $b_{kl}$ 
by soft-thresholding:
\begin{equation}
S_s\left(z, c\right)=\sgn(z)\, (\abs{z}-c)_+.
\end{equation}
In order to take into account that the frame elements $\Psi_{kl}$ are not normalized,
and generally have different norms, we use element-adapted thresholds of the form $c_{kl}=\sigma \norm{\Psi_{kl}} t$ which depend on the variance of $\scal{\epsilon}{\Psi_{kl}}$ and some global parameter $t$. Equivalently, this corresponds to first normalizing the observed coefficients
$b_{kl}$ by dividing by their variance, then applying a global threshold to the normalized coefficients, and finally inverting the normalization.

The estimator of $f$ is then the plug-in estimator
\begin{equation}\wh{f}_S
= \sum_{k,l} S\left(b_{kl}, c_{kl}\right) \Psi_{kl} \label{eq:est}
= T^* S(b,c),
\end{equation}
where $S(b,c)$ denotes the vector of thresholded coefficients, and
$T^*$ is the synthesis operator of the frame as introduced in Section~\ref{sec:frames}.

To measure the performance of this estimator, we use the risk measure 
\begin{equation}Risk(\wh{f},f)=\ewt[\epsilon]{\norm{\wh{f}-f}^2},\end{equation}
 that is, the expected quadratic norm at the sampled points  (where $\norm{f}^2=\sum_{i=1}^n f(x_i)^2$ is the Euclidean vector norm of $f$ on the observation points), 
for the performance analysis of an estimator $\wh{f}\in\RZ^n$.  

For bounding the risk of the thresholding estimator $\wh{f}_S$, 
rather than assuming some specific
regularity properties on the function $f$, it is useful to compare 
the performance of $\wh{f}_S$ to that of a group of
reference estimators. This is called the {\em oracle } approach
\cite{Candes06,Donoho94}: can the proposed estimator have a performance (almost) as good
as the best estimator (for this specific $f$) in a reference family (that is to say,
as good as if an oracle would have given us advance knowledge of which reference 
estimator is the best for this function $f$). 
We review here briefly some important results.

A suitable class of simple reference estimators consists of ``keep or kill'' (or
diagonal projection) estimators,
that keep without changes the observed coefficients $b_{k,l}$ for $(k,l)$ in some subset $I$,
and put to zero the coefficients for indices outside of $I$:
\begin{equation}
\wh{f}_I := \sum_{(k,l) \in I} b_{kl} \Psi_{kl} = T^* \wh{a}^I_{kl},
\end{equation}
where $\wh{a}^I_{kl}=b_{kl} \ind {(k,l)\in I}$.
Now using 
the frame reconstruction formula and \eqref{eq:analnorm}, we obtain
\begin{eqnarray}
\ewt[\epsilon]{\norm{\wh{f}_I-f}^2}
& = &\ewt[\epsilon]{ \norm{T^*(a - \wh{a}^I)}^2}\nonumber\\ 
&\leq &\ewt[\epsilon]{ \norm{a-\wh{a}^I}^2}\nonumber\\ 
& = &\sum_{(k,l)} \big( a_{kl}^2 \ind{(k,l)\not\in I}\nonumber\\
& &\;\;\;  + \sigma^2 \norm{\psi_{kl}}^2  \ind{(k,l) \in I}\big)\,.
\end{eqnarray}

Therefore, the optimal (oracle) choice of the index set $I^*$ obtained by minimizing
the above upper bound is given by
\begin{eqnarray}
&(k,l)\in I^*  ~~&\Leftrightarrow ~~\scal{f}{\Psi_{kl}}^2  \geq \sigma^2 \norm{\Psi_{kl}}^2   ~\mbox{~~(keep)~~}\nonumber\\
&(k,l)\notin I^*~~&\Leftrightarrow ~~\scal{f}{\Psi_{kl}}^2 \leq \sigma^2 \norm{\Psi_{kl}}^2	~\mbox{~~(kill)~~}. 
\end{eqnarray} 
One deduces from this that
\begin{equation}
\inf_I\ewt[\epsilon]{\norm{\wh{f}_I-f}^2} \leq \sum_{(k,l)\in N} \min\left(\scal{f}{\Psi_{kl}}^2, \sigma^2\norm{\Psi_{kl}}^2\right) =: OB(f)\,.
\end{equation}

The relation of soft thresholding estimators to the collection of keep-or-kill estimators on a 
Parseval frame is captured by the following oracle-type inequality 
(see \cite{Candes06}, Section 9)\footnote{\cite{Candes06} only hints at the proof;
we provide a proof in the appendix for completeness.}:

\begin{theorem}\label{s:framerisk}
Let $\{\Psi_{kl}\}_{k,l}$ be a Parseval frame and consider the denoising observation model.
Let $\wh{f}_{S_s}=\sum_{k,l} S_s\left(\scal{y}{\Psi_{kl}},t_{kl}\right) \Psi_{kl}$ be the soft-threshold frame estimator from (\ref{eq:est}). Then with $t_{kl}=\sigma\norm{\Psi_{kl}}\sqrt{2 \log(n)}$ the 
following inequality holds:
\begin{equation}
\ewt[\epsilon]{\norm{\wh{f}_{S_s}-f}^2} \leq \left(2\log(n)+1\right) \left( \sigma^2 +OB(f)\right).
\end{equation}
\end{theorem}
To interpret this result, observe that if we renormalize the squared norm by $\frac{1}{n}$,
so that it represents averaged squared error per point, we expect (depending on the
regularity of $f$) the order of magnitude of $n^{-1}OB(f)$ to be typically a polynomial rate
$O(n^{-\nu})$ for some $\nu<1$. Then the term $\sigma^2/n$ is negligible in comparison, and the
oracle inequality states that the performance of $\wh{f}_{S_s}$ is only worse by
a logarithmic factor than the performance obtained with the optimal, $f$-dependent choice
of $I$ in a keep-or-kill estimator.

For this tight oracle inequality to hold, it is particularly important that a Parseval frame is used.
While thresholding strategies can also be applied to the coefficients of a frame
that is not Parseval, the reconstruction step is less straightforward, 
(the canonical dual frame must be computed for reconstruction from
the thresholded coefficients, see
Section~\ref{sec:frames}); furthermore, an additional factor $B/A$ comes into the bound ($A\leq 1 \leq B$
being the frame bounds from definition \eqref{frameCond}) (see for instance, 
\cite{Haltmeier12}, Prop. 3.10). Therefore,
the performance of simple thresholding estimates deteriorates when used with a non-Parseval frame.

\section{Numerical experiments}

\begin{figure*}
\centering
\includegraphics[width=0.4\textwidth]{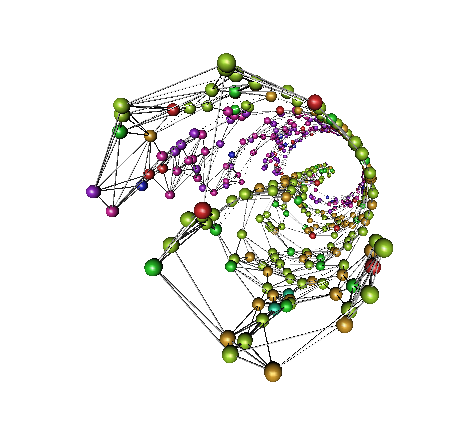}
\includegraphics[width=0.586\textwidth]{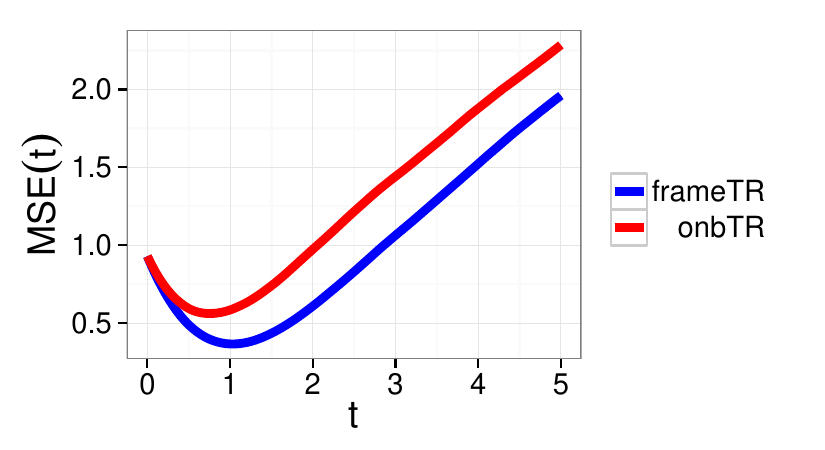}\\
\includegraphics[width=0.4\textwidth]{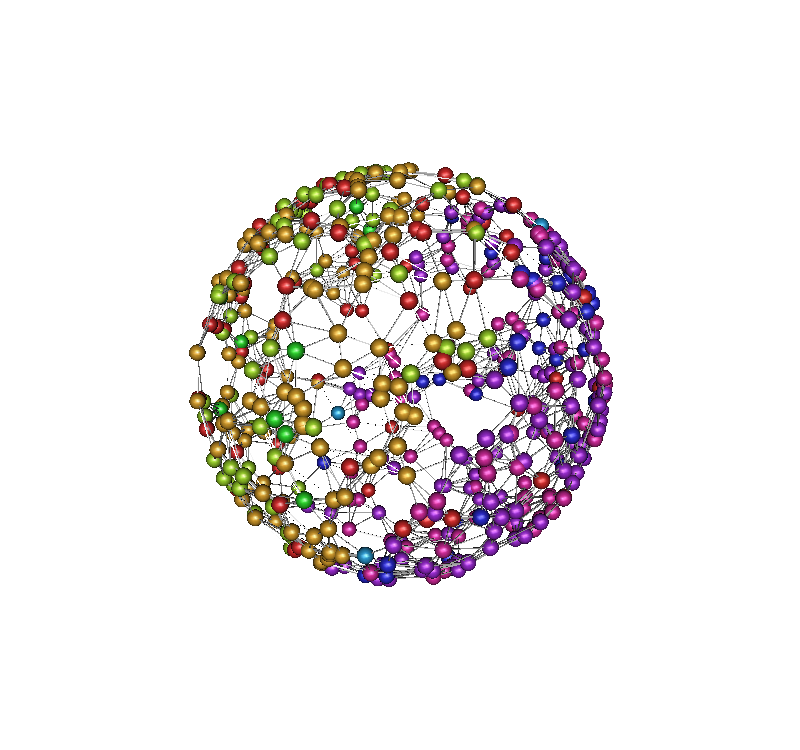}
\includegraphics[width=0.586\textwidth]{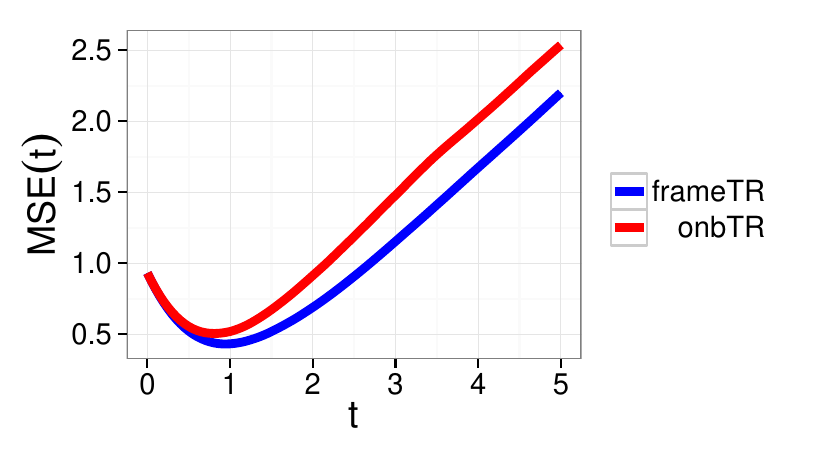}

\caption{\label{fig:exp}Left: noisy function on swiss roll data (top) and sphere data (bottom), graph representation.
Right: MSE for two representative settings (weighted $\eps$-Graph and $k$-NN-Graph) as a function of threshold level.
Red is thresholding in the original Laplacian Eigenmaps ONB, blue is thresholding
of frame coefficients.}
\end{figure*}

\label{sec:exp}

 \setlength{\tabcolsep}{0.5cm}
\begin{table}
\caption{\label{MSEtable} MSE performance %over different settings and methods
under optimal parameter choice. FrTh = Frame Tresholding; LETh/LETr = Laplacian Eigenmaps
Thresholding/Truncated expansion. Prefix W indicates edge weighting in the graph.
CGK is the complete graph with Gaussian weights. U/N is un/normalized graph Laplacian.Standard error in brackets.Top: Sphere example. Bottom: swiss roll example}
\label{table1}
\centering

%%%%%%results from experiments.pdf (Oleksandr) vom 28.9.16
Example:  sphere, jump function,$\sigma^2=1, n=500, m=50$
\rowcolors{3}{gray!20}{white}
\begin{tabular}[c]{l l r r r}
\textbf{Graph} & \textbf{$L$ }& {\textbf{FrTh}} & {\textbf{LETh}}& {\textbf{LETr}}\\
\hline 
kNN & U &		0.510 $~$ (0.050) 	& 0.693 $~$ (0.061)	& 0.905 $~$ (0.108)\\
kNN &N& 		0.538 $~$ (0.046) 	& 0.712 $~$ (0.055)	& 0.931 $~$ (0.094)\\
WkNN &U& 		0.521 $~$ (0.049)	& 0.652 $~$ (0.050) 	& 0.800 $~$ (0.097)\\
WkNN &N& 		0.530 $~$ (0.049) 	& 0.674 $~$ (0.057) 	& 0.749 $~$ (0.091)\\
CGK& U& 		0.520 $~$ (0.055) 	& 0.638 $~$ (0.065) 	& 0.821 $~$ (0.107)\\
CGK &N& 		0.530 $~$ (0.052) 	& 0.670 $~$ (0.050) 	& 0.725 $~$ (0.081)\\
$\epsilon$G &U& 	0.505 $~$ (0.058) 	& 0.650 $~$ (0.068) 	& 0.865 $~$ (0.115)\\
$\epsilon$G& N& 	0.557 $~$ (0.052) 	& 0.710 $~$ (0.059) 	& 0.902 $~$ (0.106)\\
W$\epsilon$G &U&	0.482 $~$ (0.055) 	& 0.622 $~$ (0.064) 	& 0.787 $~$ (0.111)\\
W$\epsilon$G &N& 	0.530 $~$ (0.049) 	& 0.674 $~$ (0.057) 	& 0.749 $~$ (0.091)\\
\hline
\end{tabular}
\begin{center}
 Smoothing Kernel Regression: min. MSE =  0.612 (0.066)\\
 Kernel Ridge Regression: min. MSE =  0.594  (0.051) \\
\end{center}
\hspace*{1cm}

Example: swiss roll, jump function,$\sigma^2=1, n=500, m=50$
\rowcolors{3}{gray!20}{white}
\begin{tabular}[c]{l l r r r}
\textbf{Graph} & \textbf{$L$ }& {\textbf{FrTh}} & {\textbf{LETh}}& {\textbf{LETr}}\\
\hline 
kNN & U& 		0.462 $~$ (0.043)  &	0.647 $~$ (0.039) 	& 0.876 $~$ (0.079)\\
kNN &N& 		0.494 $~$ (0.043)&	0.676 $~$ (0.043) 	& 0.902 $~$ (0.071)\\
WkNN &U& 		0.443 $~$ (0.045) &	0.600 $~$ (0.050) 	& 0.790 $~$ (0.102)\\
WkNN &N& 		0.500 $~$ (0.043)& 	0.659 $~$ (0.045) 	& 0.775 $~$ (0.079)\\
CGK &U& 		0.491 $~$ (0.053)& 	0.625 $~$ (0.057) 	& 0.844 $~$ (0.096)\\
CGK &N& 		0.520 $~$ (0.047)&	0.648 $~$ (0.049) 	& 0.713 $~$ (0.079)\\
$\epsilon$G &U& 	0.459 $~$ (0.049) &	0.610 $~$ (0.053) 	& 0.872 $~$ (0.095)\\
$\epsilon$G &N& 	0.532 $~$ (0.045) &	0.681 $~$ (0.050) 	& 0.884 $~$ (0.089)\\
W$\epsilon$G &U&	 0.441$~$  (0.049)&	0.574 $~$ (0.049)	& 0.793 $~$ (0.113)\\
W$\epsilon$G &N& 	0.503 $~$ (0.045) &	0.643 $~$ (0.051) 	& 0.744 $~$ (0.089)\\
\hline
\end{tabular}
\begin{center}
 Smoothing Kernel Regression: min. MSE = 0.589 (0.082)\\
 Kernel Ridge Regression: min. MSE = 0.779 (0.052) 
\end{center}
\end{table}

We investigate the performance of the proposed method for denoising on two testbed datasets where the
ground truth is known and the design points are drawn randomly iid from a distribution on a manifold. More precisely, we will consider one example where the design points $\data{D}$ are drawn uniformly ($n=500$) on the unit square, which is
then rolled up into a ``swiss roll'' shape in 3D. We consider a very simple target function 
represented (on the original unit square) as a piecewise constant function (with values 5 and -3) on two triangles, 
displaying a sharp discontinuity along one diagonal of the square and very smooth regularity 
elsewhere. This function is observed with an additional Gaussian noise
of variance $\sigma^2=1$.
In the second example the design points $\data{D}$ are drawn uniformly ($n=500$) on the unit sphere in $\RZ^3$. The target function remains a  piecewise constant function, defined on the two parts of the sphere when intersecting it with a chosen plane. Again, this function is observed with an additional Gaussian noise
of variance $\sigma^2=1$.  For the swiss roll example as well as for the sphere example, one sample consisting of design points and noisy function values is displayed in Figure~\ref{fig:exp}.

In each example, we consider the different types of neighborhood graphs described in
Section~\ref{sec:graph}. Following usual heuristics, for the construction of the $k$-NN graph
we take $k=7 \approx \log n$; for the $\eps$-graph, we take for $\eps$ the average distance to the
$k=7$th nearest neighbor, and for weighted graphs we take Gaussian weights, where the bandwidth 
$\lambda$ is calibrated so that points at the distance $\eps$ defined above are given weight $0.5$.

After constructing the (weighted or unweighted) graph Laplacian, we compute explicitly its eigendecomposition.
For the construction of the frame via the multiscale bandpass filter, we use a $\cC^3$ piecewise polynomial
plateau function $g$ satisfying the support constraints of Definition~\ref{bandpass} for $b=2$ (i.e. constant
equal to 1 for $x\leq 0.5$, and zero for $x\geq 1$). While this function 
is not $\cC^\infty$, it has the advantage of fast computation. 

We compare the denoising performance of the following competitors: Parseval frame with soft thresholding,
soft thresholding applied to the Laplacian Eigenmaps orthormal basis, and truncated expansion in the
Laplacian Eigenmaps basis (only the $k$ coefficients corresponding to the first eigenvalues are kept,
without thresholding). The latter method is in the spirit of \cite{LapEig2}. It is well-known (from
the regular grid case) that the ``universal'' theoretical 
threshold $\sigma\sqrt{\log n}$ is often too conservative
in practice. For a fair comparison, we therefore compute the mean squared error (MSE) 
of both thresholding methods
for varying threshold $t$ (still modulated by $\norm{\Psi_{kl}}$ for the Parseval frame).
Comparison of the MSE for one sample accross the $t$-range for two particular settings is plotted on Figure~\ref{fig:exp}.
For all studied settings (different graph and graph Laplacian types), for the same
threshold level $t$ we observed that the frame-based method systematically shows 
a noticeable improvement.

In Table \ref{table1} we report the  minimum MSEs and their standard error (averaged over $m=50$ samples of design points and independent noise)  for different methods over the possible range of the parameter
(threshold level $t$, resp. number of coefficients for truncated expansion), both for the swissroll and for the sphere example.
We observe
an improvement of 20 to 25\% across the different settings (the best overall results being obtained
with weighted graphs and the unnormalized Laplacian). We also compared to the more traditional
methods of kernel smoothing (Nadaraya-Watson estimator) and kernel ridge regression, using
a Gaussian kernel (also with optimal choices of bandwidth and regularization parameter), and
observed a comparable performance improvement. While it is not realistic to assume 
that the optimal parameter choice is known in practice, 
it is fair to compare all methods under their respective optimal parameter settings, 
as parameter selection methods will induce a comparable performance hit with respect to
the best setting.

\section{Outlook}

Following the recently introduced idea of generalizing the Littlewood-Paley spectral decomposition,
we constructed explicitly a Parseval frame
of functions on a neighborhood graph formed on the data points. We established that 
a thresholding strategy on the frame coefficients  has 
superior performance for the denoising problem as compared to usual, spectral or non-spectral,
approaches. Future developments include extension of this methodology to the 
semisupervised learning setting, and a stronger theoretical basis for spatial localization.

\begin{acknowledgement}
The authors acknowledge the financial support of the german DFG, under the Research Unit FOR-1735 ``Structural Inference in
Statistics -- Adaptation and Efficiency''.
\end{acknowledgement}

\section*{Appendix}
\addcontentsline{toc}{section}{Appendix}

\section{Proof of Theorem 3}

Theorem \ref{s:framerisk} states a oracle-type inequality which captures the relation of soft thresholding estimators  $\hat{f}_{S_s}=\sum_{k,l} S_s\left(\scal{y}{\Psi_{kl}},t_{kl}\right) \Psi_{kl}$ defined in (\ref{eq:est}) to the collection
of keep-or-kill estimators on a Parseval frame.
This result is known in the literature (see \cite{Candes06}, Section 9),
but we provide a short self-contained proof for completeness,
modulo a technical result from \cite{Donoho94} for soft thresholding 
of a single one-dimensional Gaussian variable,
 which is basic for the proof of theorem \ref{s:framerisk}.
\begin{lemma}
For $0\leq \delta\leq 1/2$, $t=\sqrt{2\log(\delta^{-1})}$ and $X\sim\NV(\mu,1)$
\begin{eqnarray}\label{lem1}
\ewt[X]{\left(S_s(X,t)-\mu\right)^2}&\leq & (2\log ( \delta^{-1})+1)(\delta+\min(1,\mu^2) )\nonumber\\
&=&(t^2+1)\left( \exp \left(-\frac{t^2}{2}\right) +\min(1,\mu^2) \right).
\end{eqnarray}
\end{lemma}
The proof of this lemma can be found in appendix 1 of \cite{Donoho94}. Now we are able to prove theorem \ref{s:framerisk}.
\begin{proof} %of theorem 4.1:\\
First note that for $y=\tau x,\tau>0$, we have 
\begin{equation}\label{eq:2}
S_s(y,u)=\tau S_s\left(x, \frac{u}{\tau}\right).
\end{equation}
Secondly we remark that 
\begin{equation}
\frac{\scal{y}{\Psi_{kl}}}{\sigma\norm{\Psi_{kl}}}\sim\NV\left(\frac{a_{kl}}{\sigma\norm{\Psi_{kl}}},1\right).
\end{equation}
Considering now the risk of the soft thresholding estimator $\hat{f}_{S_s}$ we get
\begin{eqnarray}
\ew{\norm{\hat{f}_{S_s}-f}^2} &=&\ew{\norm{\sum_{k,l} \left(S\left(\scal{y}{\Psi_{kl}}, t_{kl}\right)-a_{kl}\right) \Psi_{kl}}^2}\nonumber\\
&\leq &\ew{\sum_{k,l}\left(S \left(\scal{y}{\Psi_{kl}}, t_{kl} \right)-a_{kl}\right)^2}\nonumber\\
&=&\sum_{k,l} \ew{\left( S\left(\scal{y}{\Psi_{kl}}, t_{kl}\right)-a_{kl}\right)^2}.
\end{eqnarray}
by using inequality (\ref{eq:analnorm}). By applying (\ref{eq:2}) and then (\ref{lem1}) with $t=\sqrt{2\log(n)}$ it follows that
\begin{eqnarray}
&&\ew{\norm{\hat{f}_{S_s}-f}^2} \leq \sum_{k,l} \sigma^2\norm{\Psi_{kl}}^2 \ew{\left( S\left(\frac{\scal{y}{\Psi_{kl}}}{\sigma\norm{\Psi_{kl}}}, \sqrt{2\log(n)}\right)-\frac{a_{kl}}{\sigma\norm{\Psi_{kl}}}\right)^2}\nonumber\\
&\leq &  \sum_{k,l} \sigma^2\norm{\Psi_{kl}}^2 (2\log(n)+1)\left(\exp \left(-\frac{2\log(n)}{2}\right) +\min \left(1,\frac{a_{kl}^2}{\sigma^2\norm{\Psi_{kl}}^2}\right) \right)\nonumber\\
&=& \sum_{k,l}  (2\log(n)+1)\left( \frac{1}{n}\sigma^2\norm{\Psi_{kl}}^2 +\min \left(\sigma^2\norm{\Psi_{kl}}^2,a_{kl}^2\right)\right )\nonumber\\
&=& (2\log(n)+1) \left( \frac{1}{n} \sum_{k,l}  \sigma^2\norm{\Psi_{kl}}^2 +\sum_{k,l}  \min \left(\sigma^2\norm{\Psi_{kl}}^2,a_{kl}^2\right)        \right).
\end{eqnarray}
Recalling the Parseval frame property $\sum_{k,l} \norm{\Psi_{kl}}^2=n$, we finally obtain
\begin{eqnarray}
\ew{\norm{\hat{f}_{S_s}-f}^2}&\leq & (2\log(n)+1) \left( \frac{1}{n}n \sigma^2  +\sum_{k,l}  \min \left(\sigma^2\norm{\Psi_{kl}}^2,a_{kl}^2\right)        \right)\nonumber\\
&=& (2\log(n)+1) \left( \sigma^2  +\sum_{k,l}  \min \left(\sigma^2\norm{\Psi_{kl}}^2,a_{kl}^2\right)        \right).
\end{eqnarray}
where we recognize the upper bound $\sum_{k,l}  \min \left(\sigma^2\norm{\Psi_{kl}}^2,a_{kl}^2        \right)=OB(f)$ for the oracle.
\qed
\end{proof}

%\input{referenc}
% BibTeX users please use

% \bibliographystyle{authordate1}
\bibliographystyle{spmpsci}
 \bibliography{Abibo}%

\end{document}